# TrustMAS: Trusted Communication Platform for Multi-Agent Systems

Krzysztof Szczypiorski, Igor Margasiński, Wojciech Mazurczyk,
Krzysztof Cabaj, Paweł Radziszewski

Warsaw University of Technology, Faculty of Electronics and Information
Technology, 15/19 Nowowiejska Str., 00-665 Warsaw, Poland
`{K.Szczypiorski, I.Margasinski, W.Mazurczyk,`
`K.Cabaj, P.Radziszewski}@elka.pw.edu.pl`

**Abstract.** The paper presents TrustMAS – Trusted Communication Platform for Multi-Agent Systems, which provides trust and anonymity for mobile agents. The platform includes anonymous technique based on random-walk algorithm for providing general purpose anonymous communication for agents. All agents, which take part in the proposed platform, benefit from trust and anonymity that is provided for their interactions. Moreover, in TrustMAS there are StegAgents (SA) that are able to perform various steganographic communication. To achieve that goal, SAs may use methods in different layers of TCP/IP model or specialized middleware enabling steganography that allows hidden communication through all layers of mentioned model. In TrustMAS steganographic channels are used to exchange routing tables between StegAgents. Thus all StegAgents in TrustMAS with their ability to exchange information by using hidden channels form distributed steganographic router (Stegrouter).

## 1 Introduction

In this paper, we present and evaluate a concept of *TrustMAS* - Trusted Communication Platform for Multi-Agent Systems which was initially introduced in [21]. For this purpose we have developed a *distributed steganographic router* and *steganographic routing protocol*. To evaluate the proposed concept we have analyzed: security, scalability, convergence time, and traffic overheads imposed by TrustMAS. Presented in this paper simulation results proved that proposed system is efficient.

TrustMAS is based on agents and their operations; an agent can be generally classified as stationary or mobile. The main difference between both types is that stationary agent resides only on a single platform (host that agent operates on) and mobile one is able to migrate from one host to another while preserving its data and state.

Generally, systems that utilize agents benefit from improved: fault tolerance (it is harder for intruder to interrupt communication when it is distributed), scalability and flexibility, performance, lightweight design and ability to be assigned to different tasks to perform. Moreover, systems that consist of many agents interacting with each other form MAS (Multi-Agent System). The common applications of MAS include:

- *network monitoring* (IDS/IPS systems like in [13]),
- *network management*,
- *information filtering and gathering* (e.g. Google),
- *building self-healing*, high scalable *networks* or *protection systems* (like proposed in [20]),
- *transportation, logistics* and others (e.g. graphics computer games development [9]).

Multi-Agent Systems are usually implemented on platforms which are the tools that simplify implementation of specific systems. The most popular examples of such platforms are: JADE [12], AgentBuilder [1], JACK [11], MadKit [15] or Zeus [24].

Mobile agents, which are used in TrustMAS, create dynamic environment and are able to establish ad-hoc trust relations to perform intended tasks collectively and efficiently. Particularly, challenging goals are authentication process where an identity of agent may be unknown and authorization decisions where a policy should accommodate to distributed and changing structure. Trusted cooperation in heterogeneous MAS environment requires not only trust establishment but also monitoring and adjusting existing relations. Currently, two main concepts of the trust establishment for distributed environment exist:

- *reputation-based* trust management (TM) ([5], [14]), which utilizes information aggregated by system entities to evaluate reputation of chosen entity; basically, decisions are made according to recommendations from other entities where some of them can be better than others; the most popular example of such trust management is PageRank implemented in Google,
- *credential-* (or *rule-*) *based* trust management ([3], [2]) that uses secure (e.g. cryptographically signed) statements about a chosen entity; decisions based on this TM are more reliable but require better defined semantics then reputation-based TM.

For MAS environment we propose a *distributed steganographic router* which will provide ability to create the covert channels between chosen agents (StegAgents). Paths between agents may be created with the use of any of the steganographic methods in any OSI RM (Open System Interconnection Reference Model) layer and be adjusted to the heterogeneous characteristics of a given network. This concept of a steganographic router, as stated earlier, is new in the steganography state of the art and also MAS technology seems to be very accurate to implement such router in this environment.

To develop safe and a far-reaching agent communication platform it is required to enhance routing process with anonymity. The first concept of network anonymity was Mixnet proposed by Chaum in [4]. It has become a foundation of modern anonymity systems. The concept of Mixnet *chaining with encryption* has been used in a wide range of applications such as E-mail ([6]), Web browsing [10] and general IP traffic anonymization (e.g. Tor [8]). Other solutions like e.g. Crowds [17], may be considered as simplifications of Mixnet. By means of forwarding traffic for others it is possible to provide every agents' untraceability. The origin of collaboration intent in this manner can be hidden from untrusted agents and eavesdroppers.

## 2  TrustMAS Concept and Main Components

This section is based on paper [21], where initial concept of TrustMAS was introduced in greater details. Examples of Steg-router operations and other key TrustMAS components may be also found in [21]. The following section only briefly describes proposed solution to focus later mainly on security and performance analysis.

### 2.1 Trust and Anonymity in TrustMAS

MAS gives an opportunity to build an agents' community. In such environments, like in human society, trust and anonymity become important issues as they enable agents to build and manage their relationships. Taking this into consideration we assume that there are no typical behaviors of the agents involved in the particular MAS community, all agents may exist and live their lives in their own way (we do not define agents' interests and there is no information about characteristics of exchanged messages). Additionally, because TrustMAS is focused on information hiding in MAS, we don't assume that any background traffic exists. Abovementioned assumptions are generic and allow to theoretically describe TrustMAS. In real environment, such as IP networks, a background traffic exists and will aggravate detection of the system.

Moreover, in order to minimize the uncertainty of the interactions each agent in TrustMAS must possess a certain level of trust for other agents. Agents interactions often happen in uncertain, dynamically changing and distributed environment. Trust supports agents in right decisions making, and is usually described as reliability or trustworthiness of the other communication sides. When the trust value is high, the party with which agent is operating gives more chances to succeed e.g. agents need less time to find and achieve their goals. On the contrary, when the trust value is low, the choice of the operating party is more difficult, time-consuming and provides less chances for success. In the proposed TrustMAS platform we provide trust and anonymity for each agent wishing to join it. Main trust model of TrustMAS platform is based on a specific behavior of agents – waiting for expected scenario and following a dialog process means that agents are trusted. Other trust models, not included in this work, depend mainly on application of TrustMAS and can be changed accordingly.

One of the important components in TrustMAS is an anonymous technique based on the random-walk algorithm [18]. It is used to provide anonymous communication for every agent in the MAS platform. The idea of this algorithm is as follows. If the agent wants to send a message anonymously, it sends a message (which contains a destination address) to a randomly chosen agent, which is selected based on the result of the flipping of an asymmetric coin (whether to forward the message to the next random agent or not). The coin asymmetry is described by a probability $p_f$. The proxy agent forwards the message to the next random proxy agent with the probability of $p_f$ and skips forwarding with a probability of $1-p_f$. This probabilistic forwarding assures anonymity because any agent cannot conclude if messages received in this manner are originated from their direct sender.

TrustMAS benefits from the large number of agents that are operating within it, because if many agents join TrustMAS it will be easier to hide covert communication (exchanged between secretly collaborating agents). Agents are likely to join the proposed MAS platform because they want to use both trust and anonymity services that are provided for their interactions. To benefit from these features each agent has to follow one rule: if it wants to participate in TrustMAS it is obligated to forward discovery steganographic messages according to the random-walk algorithm (which is described in Section 3.1). This may be viewed as the "cost" that agents have to "pay" in order to benefit from the trusted environment.

**2.2 Agents in TrustMAS**

In the TrustMAS we distinguish two groups of agents. One of them consists of *Ordinary Agents* (OAs) which use proposed platform to benefit from two security services it provides (trust and anonymity). Members of the second group are *Steganographic Agents* (StegAgents, SAs), that besides OAs functionality, use TrustMAS to perform a covert communication.

The following are features of agents in TrustMAS:
- OAs are not aware of the presence of SAs,
- OAs uses TrustMAS to perform overt communication e.g. for anonymous web surfing, secure instant messaging or anonymous file-sharing,
- SAs posses the same basic functionality as OAs but they are capable of exchanging steganograms through covert channels,
- each StegAgent is characterized by its address and steg-capabilities (which describe the steganographic techniques that SA can use to create a hidden channel to communicate with other SAs).
- StegAgents that are localized in TrustMAS platform act as a distributed steganographic router, by exchanging hidden data (steganograms) through covert channels but also if they rely on the end-to-end path between two SAs they are able to convert hidden data from one steganographic method to another,
- if in proposed platform malicious agents exist trying to uncover SAs (and their communication exchange), certain mechanisms are available (described in later sections) to limit potential risk of disclosure,
- StegAgents perform steganographic communication in various ways, especially by utilizing methods in different layers of the TCP/IP model. In particular, SAs may exploit other than application layer steganographic techniques by using specialized middleware enabling steganography through all layers in this model. In some cases there is a possibility to use only application layer steganography i.e. image or audio hiding methods. Hidden communication via middleware in different layers gives opportunity for SAs to establish links outside the MAS platform. Examples of techniques in different layers of the TCP/IP model that enable covert channels include: audio, video, still images, text hiding steganography, protocol (network) steganography or methods that depend on available medium e.g. on WLAN links a HICCUPS [22].

In TrustMAS possibility of utilizing cross-layer steganography has certain advantages. It provides more possibilities of exchanging hidden data and it is harder to uncover. However, building the communication paths with many different steganographic methods may introduce additional delays. Therefore, some state of the art information hiding techniques may be not sufficient to carry network traffic (reminding that in some steganographic applications delay is not the best measure, because the best one is just to be hidden).

### 2.3 TrustMAS Three-plane Architecture

The proposed architecture of the TrustMAS may be described on three planes (Fig. 1). In the *MAS PLATFORMS* plane, the gray areas represent homogenous MAS platforms, black dots represent StegAgents and white ones - Ordinary Agents involved in TrustMAS. StegAgents act as a distributed steganographic router (Steg-Router) as shown on *STEG ROUTING* plane. Connections are possible between StegAgents with the use of hidden channels, located in different network layers (*NETWORK* plane), and at the platform level. As mentioned earlier, the choice of steganographic methods used to communicate between each StegAgents, depends on their steg-capabilities.

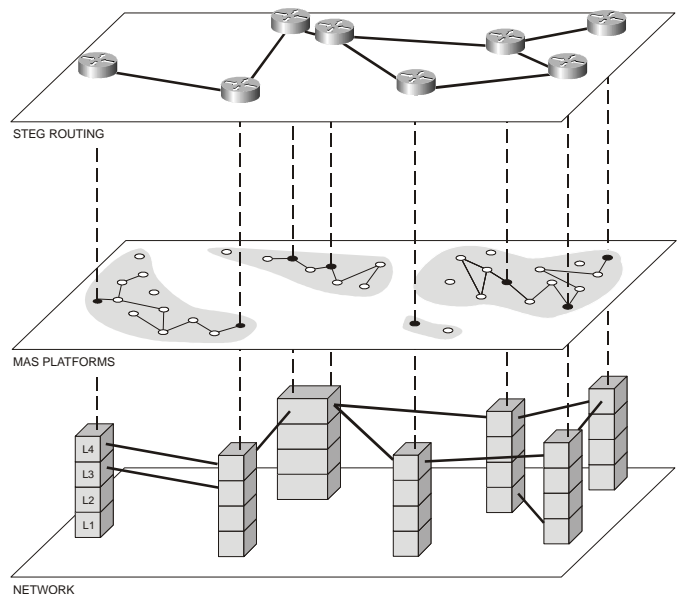

**Fig. 1** Architecture of TrustMAS

## 3  Steg-router: Distributed Steganographic Router

As mentioned earlier, all StegAgents in TrustMAS with their ability to exchange information by using hidden channels form a distributed steganographic router (Steg-router). Proposed Steg-router is a new concept of building a distributed router to carry/convert hidden data through different types of covert channels, where typically a covert channel utilizes only one steganographic method and is bounded to end-to-end connection. Moreover, it is responsible for creating and maintaining the covert channels (steg-paths) between chosen SAs. Conversion of hidden channels is performed in heterogeneous environment (e.g. a hidden information in an image converted into the hidden information in WLAN) and the MAS platform is used here as the environment to implement this concept. This gives opportunity to evaluate a new communication method and explore new potential threats in the MAS environment.

The most important part of the proposed Steg-router is a *steganographic routing protocol* (Steg-routing protocol) which is described in next sections. The effective routing protocol is vital for agents' communication and their performance. The routing protocol that will be developed for TrustMAS must take into account all specific features that cannot be found in any other routing environment. That includes providing anonymity with the random walk algorithm (and to perform discovery of new SAs) and usage of steganographic methods. Both these aspects affect performance of the routing convergence. The first one influences updates: in order to provide anonymity service they must be periodic. The second one affects available bandwidth of the links. Due to these characteristic features the steganographic routing protocol for TrustMAS must be designed carefully. For abovementioned reasons none of the existing routing protocols for MANETs (Mobile Ad-hoc Networks) is appropriate. In agents environment, for security reasons, as well as for memory and computation power requirements, provided routing protocol is kept as simple as possible that is why should it belong to a distance vector routing protocols group. We chose a distance vector routing protocol without triggered updates for security reasons – mainly to avoid potential attacks connected with monitoring agents behavior. We can imagine a situation in which the aim of the malicious attacks is to observe agents behavior after removing a random agent from the TrustMAS. If the removed agent was a StegAgent and if the Steg-routing protocol used triggered updates then suddenly there will be a vast activity in the TrustMAS, because triggered updates will be sent to announce changes in the network topology. From the same reason a distance vector protocol was chosen over the link state or a hybrid one.

Proposed steg-routing protocol will be characterized by describing: discovery and maintenance of the neighbors (section 3.1), exchanging the routing tables (section 3.2) and creating steg-links and steg-paths (section 3.3).

### 3.1 Discovery of New SAs and Neighbors Table Maintenance

As mentioned earlier, all the agents involved in the TrustMAS (both OAs and SAs) perform anonymous exchange based on random-walk algorithm. Thus StegAgents may also utilize this procedure to send anonymous messages with embedded stegmessage (covert data), which consist of StegAgents' addresses and steg-capabilities (available steganographic methods to be used for hidden communication). Such mechanism is analogous to sending hello packets to the neighbors in classical distance vector protocols, where it is responsible for discovery and maintenance of the neighbors table. In the proposed routing protocol random walk algorithm performs only discovery role. The maintenance phase is performed by all SAs that are already involved in TrustMAS and by new SAs that want to join it.

Moreover, each StegAgent maintains in its memory two tables: neighbors and routing table. The neighbors table is created based on information obtained from random-walk algorithm operations. The neighbor relation is formed between two StegAgents if there is a steg-link (exists a covert channel – a connection using steganographic method that two SAs share) that connects them. Maintenance of the actual information in the neighbors table is achieved by sending, periodically, hello packets through formed steg-links. Such solution allows to identify the situation when one of the StegAgents becomes unavailable.

Based on the information collected from the neighbors the routing table for each SA is formed. Analogously like in standard routing protocols, the routing table possesses best steg-paths (collections of steg-link between each two SAs).

### 3.2 Routing Tables Exchange

To exchange routing tables between StegAgents steganographic channels are used. In TrustMAS routing updates are sent at regular intervals to finally achieve proactive hidden routing. Routing proactivity provides unlinkability of the steganographic connections and discovery processes. This procedure as well as further hidden communication is cryptographically independent. After the discovery phase, when the new SA's neighbors table has the actual information, it receives entire routing tables from its neighboring StegAgents. Then the routing information is exchanged periodically between SAs. When a new SA receives the routing tables from its neighbors, it learns about other distant SAs and how to reach them. Based on this information formation of new steg-links with other SAs is possible.

If one of the SAs becomes unavailable, the change is detected with the hello mechanism. Then the routing table is updated and the change is sent to all the neighbors in the neighbors table (when there is periodic time to send the entire routing table). Each routing entry in the routing table represents the best available steg-path to distance StegAgent with its metric. The metric is based on three factors: *available capacity* of the steg-links along the end-to-end steg-path, *delays introduced* along the steg-path and *available steganographic methods*. For security reasons some steganographic methods may be preferred over others (e.g. because they are more immune to steganalysis or less affect the content that is used to send covert data).

### 3.3 Forming Steg-links and Steg-paths

*Steg-path* is an end-to-end connection between two distant StegAgents. Every steg-path is created based on available steg-links between SAs that form the steg-path. The algorithm of forming a steg-path uses metrics that are set for each steg-link. Routing metrics in TrustMAS are calculated as described in section 3.2.

In case there are two equal hops to one destination available, the chosen steg-link is the one that has higher capacity value, introduces less delay and uses more preferred steganographic method. It is also possible that on the one steg-link two or more steganographic methods may be available. In this case metrics are calculated for each steganographic method and the best is chosen to the steg-path. Each SA is also responsible, if it is necessary, for converting steganographic channels according to the next hop SA steg-capabilities. In this way a steganographic router functionality is provided in TrustMAS.

If the routing table is created and up to date then StegAgent is able to send data via hidden channels, where metrics are calculated based on the available steganographic methods.

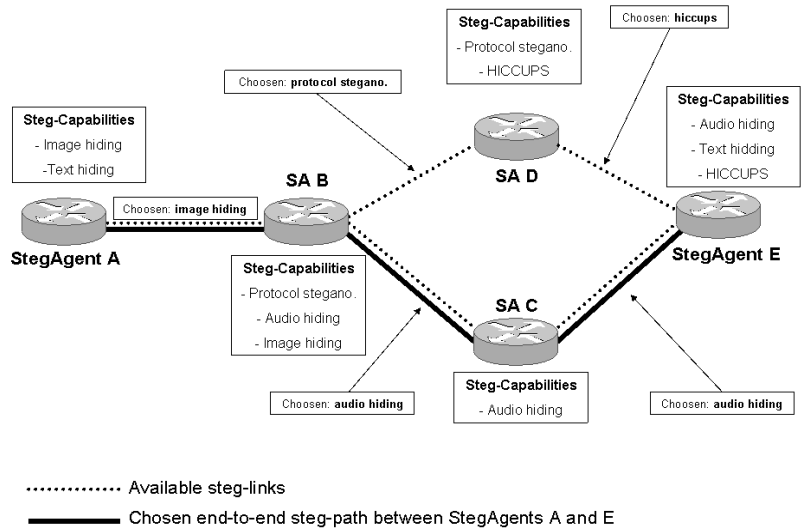

**Fig. 2** Forming a steg-path based on available steg-links between SAs

Fig. 2 illustrates a simple example for six StegAgents between which five exemplary steg-links are created based on their steganographic capabilities. The discovery phase of the the StegAgents is omitted. Based on all available steg-links an end-to-end steg-path is formed between StegAgent A and E (through proxy SAs B and C). Created steg-path consists of three steg-links. For each steg-link there is a steganographic method selected which will be used between neighboring StegAgents for hidden communication. Every proxy StegAgent that relays covert exchange is responsible for conversion of hidden data between steganographic methods that it supports (e.g. in Fig. 2 if hidden data is sent through an end-to-end steg-path SA B is obligated to convert a steganogram from image to audio steganography).

## 4 TrustMAS Security Analysis

Security analysis of TrustMAS will cover an analytical study of protocol based on an entropy measurement model. In 2002 Diaz et al. [7] and Serjantov et al. [18], simultaneously and independently, introduced a new methodology for anonymity measurement based on Shannon's information theory [19]. The information entropy proposed by Shannon can be applied to the anonymity quantification by assignment of probability of being an initiator of a specified action in the system to its particular users, nodes or agents.

The adversary who foists colluding agents on the network can assign probabilities of being the initiator to particular agents. Based on [7] and [18] we can assign such a probability to the predecessor of the first colluding agent from the forwarding path

$$p_{c+1} = 1 - p_f \frac{N-C-1}{N}. \qquad (1)$$

The rest of the agents will be assigned equal probabilities as the adversary has no additional information about them. All colluding agents should not be considered.

$$p_i = \frac{p_f}{N}, \qquad (2)$$

then

$$H_{paTM} = \frac{N - p_f(N-C-1)}{N} \log_2\left(\frac{N}{N - p_f(N-C-1)}\right) + \frac{p_f}{N}(N-C-1)\log_2\left(\frac{N}{p_f}\right). \qquad (3)$$

In the analyzed scenario it is assumed that the adversary has yet colluding agents among nodes which actively anonymize specified request. Practically, the scenario may be different, and what is more, a probability that the adversary can find this group of agents (referred to as an "active set") also determines the efficiency of the system anonymization [16]. The scenario described above should be called adaptive attack as it is assumed that the adversary has possibilities to adapt an area of his observation to the scope of activity of system users. Though it is important to consider also more general case where the adversary cannot be certain of successive collaboration of proper active agents. This attack will be referred to as a static attack as in this scenario the adversary "injects" colluding agents in a static manner and cannot dynamically predict which random agents will actively anonymize the specified request. A probability that none of the collaborating agents can become a member of the random-walk forwarding path is

$$p_r = \frac{N-C}{N}(1-p_f)\sum_{i=0}^{\infty}\left(\frac{N-C}{N}p_f\right)^i = 1 - \frac{C}{N - p_f(N-C)}, \qquad (4)$$

then an entropy for passive-static attacks equals

$$H_{psTM} = -\frac{C}{N - pf(N-C)} \frac{N - p_f(N-C-1)}{N} \log_2\left(\frac{N - p_f(N-C-1)}{N}\right) + \\ \left(1 - \frac{C}{N - pf(N-C)}\right) p_f \frac{N-C-1}{N} \log_2\left(\frac{p_f}{N}\left(1 - \frac{C}{N - pf(N-C)}\right)\right). \qquad (5)$$

Figure 2 shows entropy of TrustMAS as a function of the parameter number of colluding agents $C$ for both adaptive and static attacks.

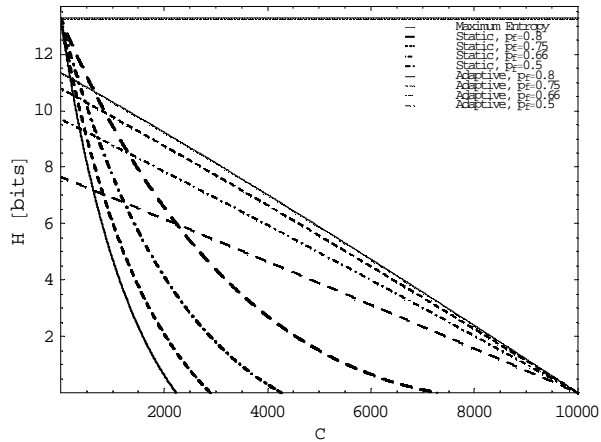

**Fig. 2.** Impact of the number of collaborating agents *C* on Entropy of TrustMAS, Static and Adaptive Attacks, $N = 10 \times 10^3$.

As one can expect, the entropy highly depends on the number of colluding agents. What is more, we can observe a significant impact of the static observation for the anonymity of TrustMAS system. TrustMAS entropy is significantly lower for static attacks than for adaptive scenarios.

Next we will analyze how exactly $p_f$ configuration impacts the entropy of TrustMAS system for both attack scenarios. Figures 3 and 4 show entropy of TrustMAS in the full spectrum of available $p_f$ configuration.

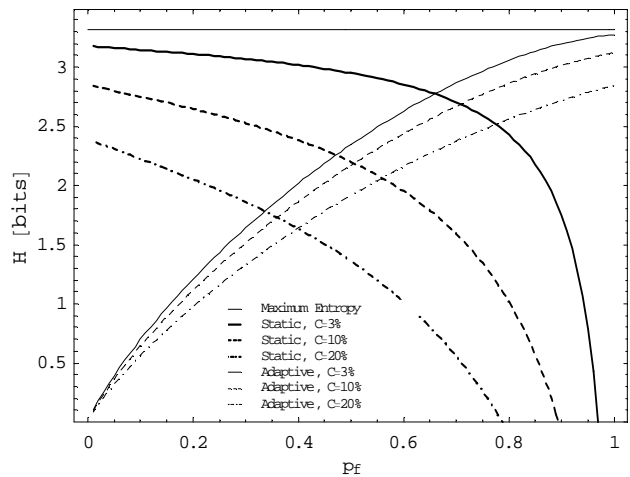

**Fig. 3.** Entropy of TrustMAS, Static and Adaptive Attacks, *N*=10.

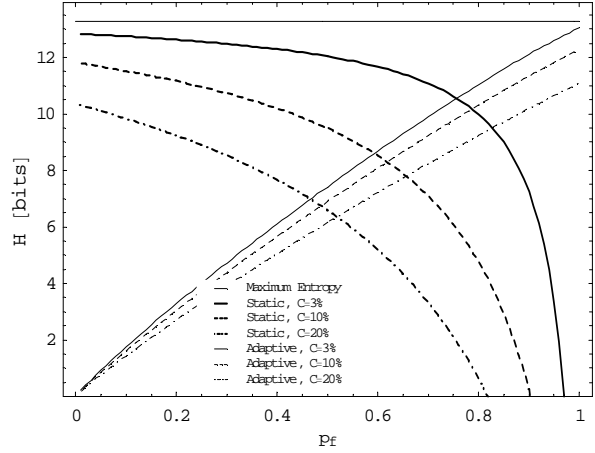

**Fig. 4.** Entropy of TrustMAS, Static and Adaptive Attacks, $N = 10 \times 10^3$.

In the adaptive scenario, a low entropy (close to zero) is obtained for low $p_f$ values and a high (close to the maximum) entropy is achieved for large $p_f$. In the static scenario, the dependency is quite different and the best results are achieved for the lowest $p_f$ values. As $p_f$ grows, the entropy grows slower logarithmically. This decrease (the static attack) of entropy is slightly faster in the small network, contrary to the adaptive scenario, where, in the small network, the decrease of entropy is slower than for large agent platform. Longer cascades can impose not only larger traffic overheads but can also make it easier for the adversary to become a member of this set and effectively compromise the security of particular systems, especially when we consider small networks. In a small network, agents from the forwarding path constitute a significant part of all network nodes.

The results show that $p_f$ configuration of TrustMAS should be in the range of [0.66 .. 0.8]. Values lower than 0.66 expose the originator against the adaptive adversary and values higher than 0.8 compromise him by the static attacker. Mean random-walk path length is

$$P = \sum_{i=2}^{\infty} i p_f^{i-2}(1-p_f) = \frac{p_f - 2}{p_f - 1}, \qquad (6)$$

then we can stress that acceptable TrustMAS mean path lengths are:
- minimum: $P_{\min TM} = 4$ for $p_f = 0.66$,
- maximum: $P_{\max TM} = 6$ for $p_f = 0.8$.

Forwarding paths shorter than $P_{\min TM}$ cannot provide sufficient "crowd" of agents which actively anonymize the initiator. If the adversary is yet among this set of agents there should be additional 3 other honest agents. On the other hand, the forwarding paths longer than 6 agents ($P_{\max TM}$) become too easy to enter as the quantity of "crowd" provided by agents that passively anonymize the active set becomes insufficient.

## 5   Traffic Performance

Based on the results achieved during the security evaluation we have analyzed TrustMAS traffic performance. Our goal was to measure convergence efficiency of proposed routing protocol and its overheads. We have designed and developed own MAS simulation environment (written in C++) that allowed us to evaluate: level of known routes among SAs, traffic generated by routing protocol for SAs, usage of platform's capacity, and SAs' links saturation levels. Presented results have been achieved under the following assumptions:

1. Simulation time $T = 30$ min. – after this period we have observed the stable operation of the system.
2. Number of agents $N \in \{250, 500, 1000, 5000, 10000\}$ – includes small and large sizes of agent community.
3. StegAgents percentage $N_{SA} = 10\%$ – typical for open and distributed network environments top limit of agents level controlled by one entity.
4. Probability $p_f \in \{0.66, 0.75, 0.8\}$ – obtained during the security analysis of the TrustMAS.
5. Migration rate $M \in \{0, 120^{-1}, 60^{-1}\}$ $[s^{-1}]$ – during traffic performance analysis we have observed that from $M = 60^{-1}$ TrustMAS operates unstable.
6. We selected six generic steganographic methods:
    - Network (Internet), bandwidth: 300000 delay: 0, probability: 0.90
    - Image, bandwidth: 100, delay: 0, probability: 0.10
    - Video, bandwidth: 100, delay: 0, probability: 0.10
    - Audio, bandwidth: 80, delay: 0, probability: 0.10
    - Text, bandwidth: 80, delay: 0, probability: 0.05
    - Network (HICCUPS), bandwidth: 225000, delay: 0, probability: 0.05
7. Routing timers were chosen based on EIGRP routing protocol defaults. Default values from EIGRP were chosen because it is one of the most efficient distance vector routing protocols.

First steganographic group describes all techniques that involves protocol steganography for Internet network. That includes e.g. IP, UDP/TCP, HTTP, ICMP etc. steganography. Because of these protocols popularity and the amount of traffic they generate, we assumed covert bandwidth's value for this steganographic group at 300 kbit/s and probability of occurrence for StegAgents in TrustMAS platform at 0.9. Next, there are four steganographic groups that correspond to techniques for data hiding in the digital content that may be sent through the network (voice, image, video and text respectively). We assumed a covert bandwidth for these steganographic methods from 80 to 100 bits/s and probability of occurrence for a StegAgent between 0.05 and 0.1. The last steganographic group characterizes more rarely used steganographic methods, e.g. medium-dependant solutions like HICCUPS. As stated earlier, the achieved steganographic bandwidth for this method may be, in certain conditions, about 225 kbit/s and this value was used during simulations.

### 5.1 Convergence Analysis

We consider the mean convergence level characteristics under dynamically changing network traffic conditions. First we analyze system behavior for no-migration scenario and then for scenarios with migration rates $M = 120^{-1}$ min$^{-1}$ and $M = 60^{-1}$ min$^{-1}$ respectively. Simulation results show 95% confidence intervals and from 25% to 75% quantiles surrounding the mean levels of known routes.

The full convergence is achieved after about 9 minutes in no-migration scenario, under dynamically changing conditions, for $M=60^{-1}$ the TrustMAS platform is not fully converged.

In the first analyzed scenario the convergence is always achieved – confidence intervals equal the mean value (100% after about 9 minutes). In the second scenario 100% convergence level is possible, however we have observed that mean value does not reach the optimum. In the last scenario the 100% level of convergence is rarely observed.

As one expects, the higher $p_f$ values are in favor of increasing the convergence time. However all analyzed configurations provide similar results.

### 5.2 Traffic Overheads Analysis

We have observed that traffic overheads imposed by the steg-routing protocol are between 10 and 12 kbps. Lower values have been obtained for higher migration rate, as when agents leave the platform, the number of exchanged large routing tables diminishes. Similarly to the convergence analysis, we have found that $p_f$ configuration has no significant impact.

The analysis of the TrustMAS capacity usage shows that the steg-routing protocol consumes less than 0.01% of the whole platform bandwidth of steg-links.

The fraction of saturated steg-links in the observed system configuration is negligible. Even for pessimistic high migration rate the saturation of the system is close to zero.

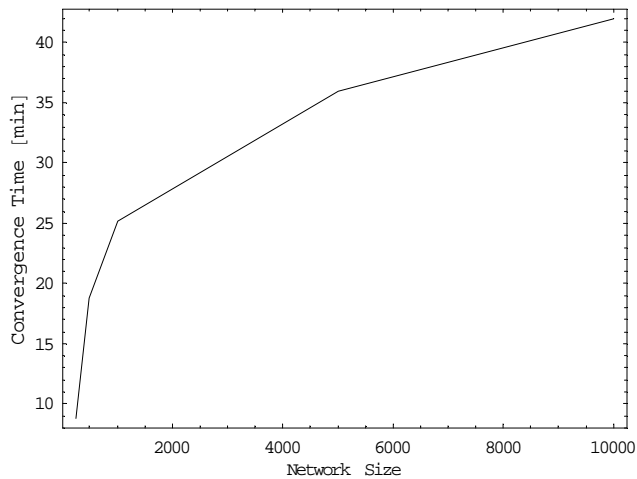

**Fig. 5** Convergence Time of TrustMAS

### 5.3 Scalability Analysis

We have repeated the traffic performance analysis for different sizes of simulated platform. We have found that the network size extends time of convergence process (Fig. 5).

Moreover, we have observed that under the stable operation of large networks ($N \in \{5000, 10000\}$), some insignificant number of undiscovered routes remains. Table 1 contains a summary of the results obtained throughout the analyzed network sizes. There, we can observe how long it takes for TrustMAS to reach a stable operation and its conditions.

**Table 1** Convergence of TrustMAS

| Network size | 250 | 500 | 1 000 | 5 000 | 10 000 |
|---|---|---|---|---|---|
| Convergence time [min] | 8.8 | 18.8 | 25.2 | 36 | 42 |
| Undiscovered routes [%] | 0 | 0 | 0 | 0,52 | 0,8 |

When we consider small networks, with 25 StegAgents collaborating among other 225 agent, less than 10 minutes is required for the proposed StegRouting protocol to provide 100% of routes between SAs. Considering very large platforms, when we have 1000 StegAgents among other 99000 agents, it would take about 40 minutes for TrustMAS to reach the stable operation. However, about 0.8% of routes would remain undiscovered.

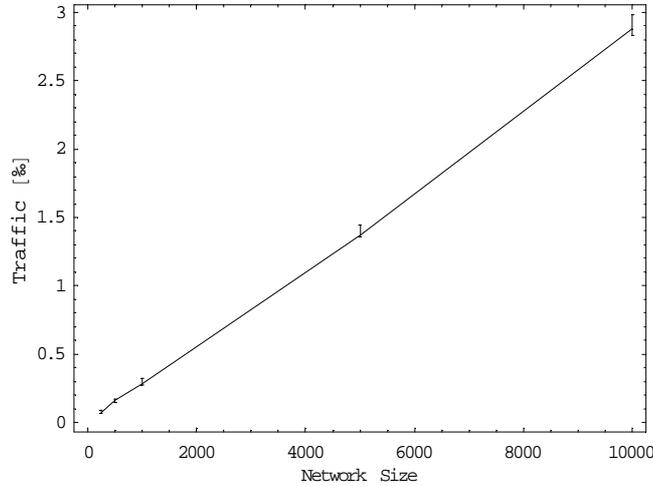

**Fig. 6** Capacity Usage of TrustMAS

The impact of the network size on TrustMAS overheads has been shown on Fig. 6. We have found that dependency between scale and the usage of the network's available bandwidth is linear. In the whole analyzed spectrum of network sizes the level of StegRouting protocol is very low, as even in the very large platform ($N = 10000$) routing management communication consumes less than 3 ‰ of all available system capacity.

The traffic performance analysis has also covered observations of the number of saturated links. In the analyzed scenario the level of links saturated by TrustMAS routing is insignificant. The highest values have been observed for the network size of $N = 1000$ agents (Fig. 7). Further extension of network size is in favor of TrustMAS communication.

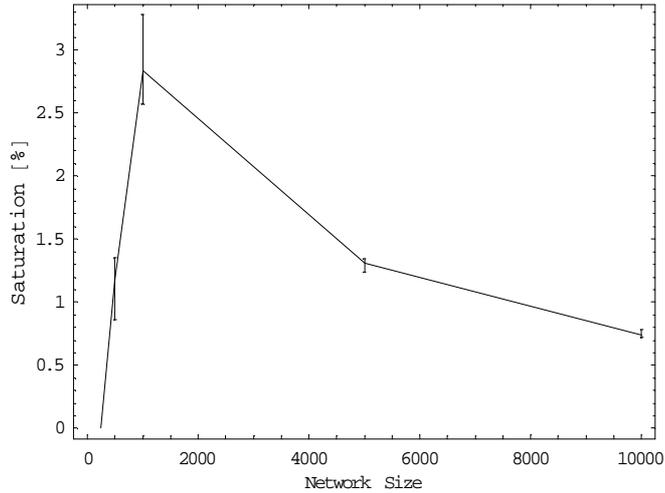

**Fig. 7** Mean Level of Saturated Links for TrustMAS

## 5 Conclusions

We have evaluated efficiency of TrustMAS both for its security and routing performance. Moreover, we have measured overheads imposed by TrustMAS platform required to assure the proper level of agents anonymity and their full connectivity. We have found that the protocol is efficient and the impact of its overheads is not significant.

Steganographic agents in the TrustMAS platform can communicate anonymously in configuration of random-walk algorithm limited to $p_f \in [0.66 .. 0.8]$. This range corresponds to a mean length of forwarding paths from $P \in [4 .. 6]$. Then, each StegAgent should involve about 3 to 5 other agents into the process of the discovery message forwarding to effectively hide an association between its identity and the sent content. Basically, at least 3 additional TrustMAS agents should relay discovery communication to hide the information that the agent is a StegAgent. On the other hand, involving more than 5 agents into the random-walk forwarding process also significantly reduces anonymity of TrustMAS. When we consider platform containing 10-20% colluding agents longer forwarding paths finally facilitate dishonest agents to penetrate the platform area where the StegAgent is hidden.

Using obtained results we have simulated the TrustMAS routing in the configuration of random-walk algorithm with $p_f \in \{0.66, 0.75, 0.8\}$ and the routing timers configuration typical for the popular EIGRP routing protocol. We have found that proposed steganographic routing is efficient and in less than 10 minutes the platform becomes fully converged. Moreover, we have observed that the protocol is robust against fast agents migration ($M = 120^{-1}$ s$^{-1}$). A borderline case was observed for high migration rate $M = 60^{-1}$ s$^{-1}$ where agents lose all discovered routing information at the average rate of one per minute.

To evaluate practical usefulness of TrustMAS we have measured the traffic overheads imposed by the proposed routing protocol. We have found that it requires about 11 kbps per link which corresponds to less than 0.01% of the system capacity. The fraction of saturated links is also negligible.

The traffic performance analysis confirmed our expectation that the impact of the random-walk $p_f$ configuration is not significant for platform overheads as the short discovery messages generate low traffic. However, higher values of $p_f$ are in favor of the routing efficiency as including 1 more agent into the discovery forwarding process provides the convergence faster by about 1 minute.

The foregoing results have been obtained for platform of hundreds of agents ($N = 250$). Taking into account a possible global and large scale environment of TrustMAS operation we have analyzed behavior of proposed protocol with simultaneous $N \in \{500, 1000, 5000, 10000\}$ communicating agents. We have found that the large scale of the network does not significantly reduce the system performance. However, in very large platforms with $N = 5000..10000$, some imperfection of the proposed routing protocol has been exposed. We should bear in mind that in such a scale the proposed distance vector protocol will discover about 99.5..99.2% of all the available routes. Still, the proposed solution scales well and can operate vastly in large scale networks.

We have proven that the proposed system is secure, fast-convergent and scalable. It can efficiently hide collaboration of designated agents (i.e. StegAgents) in various scale networks (up to ten of thousands agents). Moreover, we have proven that TrustMAS quickly enables connectivity among SAs. The convergence for small and medium size networks is fully achieved and for very large scale networks the proposed distance vector routing protocol does not discover insignificant number of routes. The overheads imposed by routing protocol are negligible.

Future work will include routing protocol improvements to support large platforms (more than 5000 agents) to eliminate negative routing effects (e.g. routing loops) and to gain faster convergence time than currently achieved. Moreover, different analyses of various scenarios for other steganographic profiles may be performed and a concept of TrustMAS may be adopted to the other environments than MAS. Additionally, a prototype of the proposed system for proof-of-concept purposes will be created and analyzed.

## Acknowledgment

This material is based upon work supported by the European Research Office of the US Army under Contract No. N62558-07-P-0042. Any opinions, findings and conclusions or recommendations expressed in this material are those of the authors and do not necessarily reflect the views of the European Research Office of the US Army.

# References

1. AgentBuilder, http://www.agentbuilder.com
2. Blaze, M., Feigenbaum J., Keromytis, A.: KeyNote: Trust Management for Public-Key Infrastructures. Springer-Verlag, LNCS, vol. 1550, 59–63 (1999)
3. Blaze, M., Feigenbaum, J., Lacy, J.: Decentralized Trust Management. In Proc. of: IEEE 17th Symposium on Research in Security and Privacy, 164–173 (1996)
4. Chaum, D.: Untraceable Electronic Mail, Return Addresses, and Digital Pseudonyms. Communications of the ACM , v. 24, n. 2, 84-88 (1981)
5. Damiani, E., Vimercati, D., Paraboschi, S., Samarati P., Violante, F.: A Reputation-based Approach for Choosing Reliable Resources in Peer-to-peer Networks. In Proc. of: The 9th ACM Conference on Computer and Communications Security CCS'02, ACM Press, Washington, DC, USA, 207–216 (2002)
6. Danezis, G., Dingledine, R., Mathewson, N.: Mixminion: Design of a Type III Anonymous Remailer Protocol. In Proc. of: The IEEE Symposium on Security and Privacy (2003)
7. Diaz, C., Seys, S., Claessens J., Preneel, B.: Towards Measuring Anonymity. In Proc. of: PET'02 - Designing Privacy Enhancing Technologies, Springer-Verlag, LNCS vol. 2482, 54–68, 2003
8. Dingledine, R., Mathewson, D., Syverson, P.: Tor: The Second Generation Onion Router. in Proceedings of the 13th USENIX Security Symposium (2004)
9. Doyle, P.: Believability through Context: Using Knowledge in the World to Create Intelligent Characters. In Proc. of: the International Joint Conference on Autonomous Agents and Multi-Agent Systems (2002)
10. Handel, T., Sandford, M.: Hiding Data in the OSI Network Model. In Proc. of: First International Workshop on Information Hiding 1996. Springer-Verlag, LNCS, vol. 1174, 23-38 (1996)
11. JACK, http://www.agent-software.com.au
12. JADE, http://jade.tilab.com
13. Jansen, W., Karygiannis, T.: NIST Special Publication 800-19 – Mobile Agent Security (2000)
14. Lee, S., Sherwood, R., Bhattacharjee, B.: Cooperative peer groups in nice. In Proc.: INFOCOM 2003. Twenty-Second Annual Joint Conference of the IEEE Computer and Communications Societies. IEEE, volume 2, 1272–1282 (2003)
15. MADKIT, http://www.madkit.org
16. Margasiński, I., Pióro, M.: A Concept of an Anonymous Direct P2P Distribution Overlay System. In proceedings of the 22nd IEEE International Conference on Advanced Information Networking and Applications (AINA2008), Okinawa, Japan (2008)
17. Reiter, M., Rubin, A.: Crowds: Anonymity for Web Transactions. ACM Transactions on Information and System Security (TISSEC), 1(1):66–92 (1998)
18. Serjantov, A., Danezis, G.: Towards an Information Theoretic Metric for Anonymity. In Proc. of: PET'02 - Designing Privacy Enhancing Technologies, Springer-Verlag, LNCS vol. 2482, 41–53 (2002)
19. Shannon, C.: A Mathematical Theory of Communication. The Bell System Technical Journal, 27:379–423:623–656 (1948)
20. Sheng, S., Li, K.K., Chan, W., Xiangjun Z., Xianzhong, D.: Agent-based Self-healing Protection System, IEEE Transactions on Volume 21, Issue 2, 610-618 (2006)
21. Szczypiorski, K., Margasiński, I., Mazurczyk, W.: Steganographic Routing in Multi Agent System Environment - Journal of Information Assurance and Security (JIAS), Dynamic Publishers Inc., Atlanta, GA 30362, USA, Volume 2, Issue 3, ISSN 1554-1010, 235-243 (2007)

22. Szczypiorski, K.: HICCUPS: Hidden Communication System for Corrupted Networks. In Proc. of: The Tenth International Multi-Conference on Advanced Computer Systems ACS'2003, Międzyzdroje, Poland, 31-40 (2003)
23. Weiss, G. (Editor): Multiagent Systems: A Modern Approach to Distributed Artificial Intelligence, Chap. 12, MIT Press, 505-534 (1999)
24. Zeus, http://www.labs.bt.com/projects/agents/zeus